\documentclass[prl,twocolumn,showpacs,superscriptaddress,nofootinbib]{revtex4-1}
\usepackage{graphicx}
\usepackage{bm}
\usepackage{amssymb}
\usepackage{color}
\usepackage{amsmath}
\usepackage{amstext}
\usepackage{latexsym}
\usepackage[colorlinks,linkcolor=blue]{hyperref}

\usepackage{mathptmx} 

\def\k{{\rm\bf k}}

\def\la{\langle}
\def\ra{\rangle}

\newcommand{\beq}{\begin{equation}}
\newcommand{\eeq}{\end{equation}}
\newcommand{\beqa}{\begin{eqnarray}}
\newcommand{\eeqa}{\end{eqnarray}}

\begin{document}
\title{Assisted  finite-rate adiabatic passage across a quantum critical point: \\Exact solution for the quantum Ising model}
\author{Adolfo del Campo}
\affiliation{Theoretical Division,  Los Alamos National Laboratory, Los Alamos, NM, USA}
\affiliation{Center for Nonlinear Studies,  Los Alamos National Laboratory, Los Alamos, NM, USA}
\author{Marek M. Rams}
\affiliation{Vienna Center for Quantum Science and Technology, Faculty of Physics, University of Vienna, Vienna, Austria}
\author{Wojciech H. Zurek}
\affiliation{Theoretical Division,  Los Alamos National Laboratory, Los Alamos, NM, USA}

\begin{abstract}

The dynamics of a  quantum phase transition is inextricably woven with the formation  of excitations, 
as a result of the critical slowing down in the neighborhood of the critical point.
We design a transitionless quantum driving through a quantum critical point that allows one to access 
the ground state of the broken-symmetry phase by a finite-rate quench of the control parameter.
The method is illustrated in the one-dimensional quantum Ising model in a transverse field.
Driving through the critical point is assisted by an auxiliary Hamiltonian, for which
 the interplay between the range of the interaction and the modes where excitations are suppressed is elucidated. 

\end{abstract}

\pacs{03.75.Kk, 67.85.-d, 03.75.-b}

\maketitle

The complexity involved in describing a generic many-body quantum system prompted Feynman to suggest 
the use of a highly controllable quantum system as a simulator of another, generally complicated, quantum system of interest \cite{QS82}.
From this perspective, interesting quantum systems are those with a large amount of entanglement and hardly tractable in classical computers \cite{CZ12}.
Quantum simulation has become an exciting field of research, which  is being developed experimentally by
exploring a variety of platforms including  ultracold atoms, trapped ions, photonic quantum systems and superconducting circuits, among others.
Simulation of many-body interacting systems 
is particularly advanced in implementations with trapped ions \cite{qsions} where the building blocks of a digital quantum simulator for both closed \cite{Lanyon11} 
and open \cite{Barreiro11} quantum systems have been demonstrated. Moreover, while early experimental efforts have been limited to somewhat
 low number of qubits, the simulation of few-hundreds of spins with variable-range spin-spin Ising-type interactions has recently been reported \cite{Britton12}.

In a continuous quantum phase transition, divergence of length and time scales across a quantum critical point (QCP) leads inevitably to non-adiabatic dynamics.
When a parameter $\lambda$ of the Hamiltonian is changed across its critical value $\lambda_c$, the energy gap between the ground and the first excited state vanishes, and adiabaticity breaks down. The Kibble-Zurek mechanism (KZM), originally developed for classical and continuous phase transitions \cite{Kibble76,Zurek96}, predicts that the resulting density of excitations obeys a power-law scaling with the quench rate. The power-law exponent is expressed using the critical exponents at equilibrium and the dimensionality of the system \cite{Dziarmaga10,Polkovnikov11}. As a result, quantum quenches are useful to characterize universal features of a system, and to shed some light on its  dynamics out of equilibrium. 

The inevitable formation of excitation is however undesirable for a wide range of applications, 
such as the preparation of novel quantum phases in quantum simulation, and adiabatic quantum computation. Suppressing excitations is also of interest to variety of operations in the laboratory, like entangling strings of atoms \cite{Dorner03}. 
This has motivated studies including the use of  the energy gap arising from the finite size of the system \cite{Murg04},  optimal non-linear passage across a QCP \cite{BP08,SSM08}, inhomogeneous quenches \cite{ZD08,CK10,DM10}, and optimal quantum control strategies \cite{OQC}. All those approaches can be regarded as strategies to exploit or engineer a spectral gap. 
Notwithstanding, there is a need for new methods to ensure adiabaticity \cite{CZ12}. In this letter, we shall  exploit recent advances in the simulation of coherent $k$-body interactions \cite{kbody,Barreiro11} and transitionless quantum driving \cite{DR03,Berry09} to explore an  alternative to quantum adiabatic protocols,  and assist a fully adiabatic finite-rate passage across a QCP.


{\it Shortcut to the adiabatic driving of a two-level system.-}
Demirplak and Rice \cite{DR03},  and Berry \cite{Berry09} have shown the possibility of implementing a transitionless quantum driving in multilevel systems.
Let us consider the Landau-Zener (LZ) transition, the simplest model supporting the KZM \cite{Damski05}, described by a Hamiltonian:
\beqa
\label{HLZ}
H_0=
\begin{pmatrix}
\lambda(t) & \Delta  \\
\Delta & -\lambda(t) 
\end{pmatrix} = \lambda(t) \sigma^z + \Delta \sigma^x,
\eeqa
where $\sigma^{x,y,z}$ are the usual Pauli matrices. The instantaneous eigenbasis reads:
\beqa
|1(\lambda)\ra&=&\sin \theta |1  (-\infty)\ra-\cos \theta |2 (-\infty) \ra,\nonumber\\
|2(\lambda)\ra&=&\cos \theta |1 (-\infty)\ra+\sin \theta |2 (-\infty) \ra, \nonumber
\eeqa
where the angle $\theta$ obeys the relations
\beqa
\cos 2 \theta=\frac{\lambda }{\sqrt{\lambda^2+\Delta^2}},
\quad \sin 2 \theta=\frac{\Delta}{\sqrt{\lambda^2+\Delta^2}}. \nonumber
\eeqa
and the energy gap is $E_2(t)-E_1(t)=2\sqrt{\Delta^2+\lambda^2}$.
Following \cite{DR03,Berry09}, 
it is found that the Hamiltonian that drives the exact evolution of the system
 along the adiabatic solution associated with the instantaneous eigenbasis $\{|n(\lambda)\ra\}$  of $H_0$ in Eq. (\ref{HLZ}) is given by $H=H_0+H_1$ with  \cite{CommentH}
\beqa
\label{H1full}
H_1&=&i \lambda'(t)  \sum_n[|\partial_\lambda n\ra\la n|-\la n|\partial_\lambda n\ra|n\ra\la n|].
\eeqa
For the LZ crossing, upon explicit calculation one finds
\beqa
H_1=-\lambda'(t) \frac{1}{2}\frac{\Delta}{\Delta^2+\lambda(t)^2}\sigma^y. 
\label{H1LZ}
\eeqa
The adiabatic solution of  $H_0$, in which the instantaneous eigenstates pick up exclusively a phase along the evolution-- the sum of the dynamical and Berry phases --  becomes the exact solution of the time-dependent Schr\"odinger equation  associated with $H=H_0+H_1$  no matter how fast the transition is crossed, i.e. how large the rate $\lambda'(t)$ is.  This approach has recently been verified in the laboratory with an effective two state model arising in a Bose-Einstein condensate in the presence of an optical lattice \cite{Morsch12}. 


{\it Models.-}
We turn now our attention to the family of $d$-dimensional free-fermion Hamiltonians:
\beqa
\mathcal{H}_0 =\sum_{\k}\psi_{\k}^\dagger \left[ \vec{a}_\k (\lambda(t))  \cdot \vec{ \sigma}_\k \right] \psi_{\k}, 
\label{models}
\eeqa
where $\vec \sigma_{\k} \equiv (\sigma_\k^x, \sigma_\k^y,\sigma_\k^z )$ denote the Pauli matrices acting on the $\k$-mode and  $\psi_{\k}^\dagger = (c_{\k,1}^\dagger,c_{\k,2}^\dagger)$ are fermionic operators. The function $\vec a_\k (\lambda) \equiv (a^x_\k (\lambda),a^y_\k (\lambda),a^z_\k (\lambda))$ is specific for the model and the sum goes over independent ${\bf k}$-modes. Such a Hamiltonian represents a variety of systems with QCP, including in particular the Ising and XY models in $d=1$ \cite{Sachdev}, as well as the Kitaev model in $d =2$ \cite{EK} and  $d=1$ \cite{1dKitaev}. As such it has been the subject of a recent series of works on defect production induced by a quantum quench \cite{Dziarmaga10}. 

We shall use it to illustrate and investigate the possibility of driving an adiabatic passage across a QCP. Let  us consider the instantaneous eigenstates of $\mathcal{H}_0$ with eigenenergies associated with the $\k$-mode
$
\varepsilon_{\k,\pm}=\pm |\vec a_\k(\lambda)| = \pm \sqrt{a_\k^x(\lambda)^2+a_\k^y(\lambda)^2+a_\k^z(\lambda)^2}. \nonumber
$
We generalize Eq. (\ref{H1LZ}) to find the modified Hamiltonian $\mathcal{H}=\mathcal{H}_0+\mathcal{H}_1$, where 
\beqa
\mathcal{H}_1 &=& \lambda'(t)\sum_{\k}\frac{1}{2 \varepsilon_\k^2} \psi_{\k}^{\dag} \left[ (\vec a_\k (\lambda) \times \partial_\lambda  \vec a_\k (\lambda)) \cdot \vec \sigma_{\k}  \right] \psi_{\k}
\label{H1free} \nonumber
\eeqa
induces the adiabatic crossing of the QCP by driving the dynamics exactly along the instantaneous eigenmodes of $\mathcal{H}_0$. 

Without further knowledge of the explicit form of the matrix elements of $\mathcal{H}_0$, its form in real space cannot be determined.
Next, we turn our attention to a specific model.


{\it The quantum Ising model in a transverse field.-}
Consider a chain of $N$ spins described by  the 1$d$ quantum Ising model in a transverse magnetic field $g$,
\beqa
\mathcal{H}_0=-\sum_{n=1}^N(\sigma_n^x\sigma_{n+1}^x+g \sigma_n^z),
\label{H_Ising}
\eeqa
a paradigmatic model to study quantum phase transitions \cite{Sachdev} of relevance to current experimental efforts in quantum simulation \cite{Monroe11}.
We assume periodic boundary conditions $\sigma_{N+1}=\sigma_{1}$ and, for simplicity, even  $N$. 
This model exhibits a quantum phase transition at $g_c = \pm 1$ between a paramagnetic phase ($|g|>1$) and ferromagnetic phase ($|g|<1$).  

The Jordan-Wigner transformation,
 $\sigma_n^z=1-2c_n^{\dag}c_n$, $(\sigma_n^x+i \sigma_n^y)=2 c_n \prod_{l<n}(1-2c_l^{\dag}c_l)$, where $c_n$ are fermionic annihilation operators, allows us to rewrite the Hamiltonian (\ref{H_Ising}) as a free fermion model.   Below, we will limit ourselves to the plus-one-parity subspace of the Hilbert space, that includes the ground state -- note that $\mathcal{H}_0$ commutes with the parity operator $P=\prod_{n=1}^N \sigma_n^z$. In fermionic representation:
\beqa
\mathcal{H}_0=\sum_{n=1}^N\left[\left(c_n - c_n^\dagger\right)\left(c^\dagger_{n+1} + c_{n+1} \right)  - g  \left( c_n c_n^{\dagger} - c_n^{\dagger} c_n \right)\right], \nonumber
\eeqa 
with anti-periodic boundary conditions $c_{N+1}=-c_1$. Using the Fourier transform $c_n=e^{-i\pi/4} \sum_k c_k e^{ikn} / \sqrt{N}$ with momenta consistent with the boundary conditions  $k \in k^+ = (\pm \pi /N, \pm 3\pi /N, \ldots, \pm(N-1)\pi /N ) $,  we can conveniently rewrite it as
\begin{eqnarray}
\mathcal{H}_0= 2\sum_{k>0} && \psi_k^\dagger \left[ \sigma_k^z (g-\cos k)+\sigma_k^x \sin k  \right] \psi_k, \nonumber
\end{eqnarray}
where the operator $\psi_k^\dagger \equiv (c_k^\dagger, c_{-k})$. In this form, it becomes apparent  that the Ising model can be decomposed into a series of independent LZ transitions, as was first realized in \cite{Dziarmaga05}. Now, we can directly use the results of the previous section and write the supplementary Hamiltonian required for an adiabatic driving across the QCP,
\beqa
\label{Hplusberry}
\mathcal{H}_1&=&-g'(t) \sum_{k>0} \frac{1}{2} \frac{ \sin k }{g^2+1-2g\cos k } \psi_k^\dagger \sigma_k^y \psi_k. \nonumber
\eeqa
This expression is expected to be highly non-local in real space.  It can be written as
\beqa
\label{Hexp}
\mathcal{H}_1=-  g'(t) \left[  \sum_{m=1}^{N/2-1} h_m(g)  \mathcal{H}_1^{[m]} +  \frac12 h_{N/2} (g)  \mathcal{H}_1^{[N/2]} \right] . 
\eeqa
The Hamiltonian $\mathcal{H}_1^{[m]}$ includes an interaction over a range $m$. Above, we have a factor of $\frac{1}{2}$ for $m=N/2$ -- which is the largest distance in the system with periodic boundary condition -- because for even N there is only one spin over the distance $N/2$ while there are two different ones over smaller distances. 
Every $\mathcal{H}_1^{[m]}$  is independent of $g$ -- all dependence on $g$ is included in coefficients $h_m(g)$ -- and reads
\beqa
\label{H1m}
 \mathcal{H}_1^{[m]} =  2 i \sum_{n=1}^N \left (c_n c_{n+m} + c_n^\dagger c_{n+m}^\dagger \right). \nonumber
 \eeqa
The coefficients $h_m(g)$ are given by the Fourier transform,
 $$h_m(g) =\frac{1}{N}\sum_k f(k)\sin(m k),$$
  of the function
\beqa
f(k)=\frac{1}{4} \frac{\sin k}{g^2+1-2g\cos k}. 
\label{fk}
\eeqa
\noindent In the limit of large $N$ we can approximate $h_m \simeq  \int_{0}^\pi f(k) \sin(m k)  dk/ \pi $, with the result
\beqa
\label{hm}
h_m= \frac{1}{8} \left \{
\begin{array}{cc} 
\begin{split}
& g^{m-1}  & {\rm for}~ |g| < 1,  \\
& g^{-m-1}& {\rm for}~  |g| > 1.
\end{split}
\end{array}
\right.  \nonumber
\eeqa
Mapping back to spins, the supplementary Hamiltonians reads
\beqa
\label{Hmplus}
\mathcal{H}^{[m]}_1=  \sum_{n=1}^N &&\left (\sigma_n^x\sigma_{n+1}^z\cdots\sigma_{n+m-1}^z\sigma_{n+m}^y  \right. \nonumber\\
&& \left. + \sigma_n^y\sigma_{n+1}^z\cdots\sigma_{n+m-1}^z\sigma_{n+m}^x \right).
\eeqa

\noindent Some comments are in order.
Firstly, since we have represented the class of Hamiltonian in Eq. (\ref{models}) -- and in particular the Ising model (\ref{H_Ising}) --
as independent LZ crossings, the supplementary Hamiltonian $\mathcal{H}_1$ allows to adiabatically drive $any$ eigenstate of the model under consideration \cite{Commentparity}. 
This means that one can further tailor the Hamiltonian $\mathcal{H}_1$ for the purpose of driving exclusively a given subset of states, e.g., the ground states.
Further considerations along that line of reasoning are beyond the scope of this letter \cite{CommentHmod}.

\noindent Secondly, coefficients $h_m(g)$ can be neatly written as $|h_m(g)| =  e^{-(m \pm 1)/\xi(g)} /8$, where $\xi(g) = |\ln(|g|)|^{-1}$ 
is the correlation length in the Ising model \cite{McCoy71}. 
It follows that $h_m(g)\sim\mathcal{O}(1)$ for distances $m$ up to the correlation length and go to zero exponentially fast for a longer interaction range. 
At the critical point this means that $\mathcal{H}_1$ is acting along the whole chain.

\noindent Finally, Hamiltonians of the form  $\mathcal{H}^{[m]}_1$ can be efficiently implemented in trapped-ion quantum simulators using  stroboscopic techniques 
\cite{kbody,Casanova12} already demonstrated in the laboratory \cite{Barreiro11}.


{\it Finite range interactions, filtering, and the KZM.-}
We next consider a linear quench of the coupling $g(t)=g_c-\upsilon t$, through the QCP at $g_c=1$, which brings the system from the paramagnetic to the ferromagnetic phase. The evolution induced by the Hamiltonian $\mathcal{H}_1$  (\ref{Hplusberry}) is adiabatic in the instantaneous eigenbasis of the Ising model (\ref{H_Ising}) \cite{Commentparity}. However the range of interaction, e.g. at the critical point, spans over the whole chain. As a result, from a practical point of view,  one might be interested in assisting the crossing  of the QCP with an approximation to $\mathcal{H}_1$  that involves interactions of restricted range. 
\begin{figure}
\begin{center}
\includegraphics{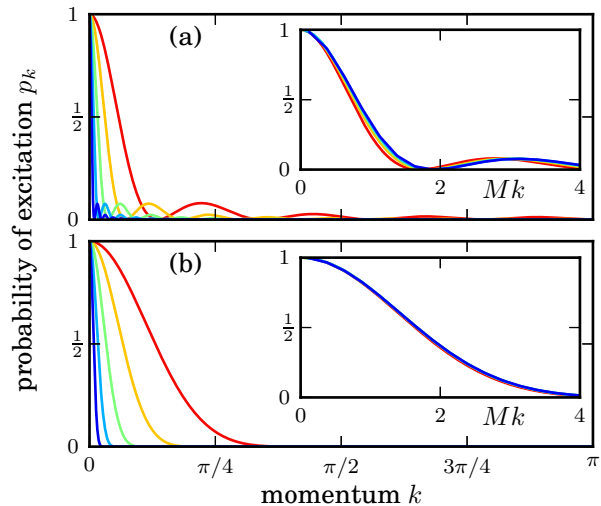}
\end{center}
\caption{\label{pkk} (Color online) Excitation probability $p_k$ as a function of the wave vector $k$, following 
a shortcut to adiabaticity in the 1$d$ quantum Ising model. (a) The crossing of the critical point 
is assisted by a truncated Hamiltonian $\mathcal{\tilde H}_1(M)$ ($s_m=1$) (b) and a modified truncation where the expansion coefficients are modulated by a raised-cosine Fourier filter.
The range of the interaction increases from right to left with cutoff $M=4,8,16,32,64$ ($N=1600$).  Above, evolve the system from $g_i=10$ to $g_f=0$ accros QCP at $g_c=1$ at quench rate $\upsilon=50$ using $\mathcal{H}=\mathcal{H}_0+\mathcal{\tilde H}_1(M)$.
The insets show the scaling of $p_k$ as function of $k/k_M\sim Mk$.
}
\end{figure}
 In this section, we examine the simplest approximation, namely, a direct truncation of Eq.  (\ref{Hexp}) that limits the range of interaction to $M$ sites,
 \beqa
 \mathcal{\tilde H}_1(M) = \upsilon \sum_{m=1}^{M} s_m h_m(g)  \mathcal{H}_1^{[m]},
 \label{H1t}
 \eeqa
where $s_m$ will be a filter function.
 
We start by examining the limit of a fast transition, which for now means $\upsilon \gg 1$  (this will be made more precise later). 
We consider the initial ground state in the paramagnetic phase and evolve it using: (i) only the supplementary Hamiltonian $\mathcal{H} = \mathcal{\tilde H}_1 (M)$, (ii) both the supplementary and the Ising Hamiltonians $\mathcal{H} = \mathcal{H}_0+ \mathcal{\tilde H}_1 (M)$. For both cases, we numerically solve the time-dependent Bogoliubov -- de Gennes equations that describe the evolution of the system, as explained in \cite{Dziarmaga05}. Fig. \ref{pkk} shows the probability of excitation in the $k$-mode, $p_k$. 

\noindent Firstly,  we have verified that $p_k$ does not depend on quench rate $\upsilon$ for $\upsilon \gg 1$, and they coincide in both cases (i) and (ii) in that limit. That is, the presence of $\mathcal{H}_0$ results only in phase difference and does not affect how well the approximated supplementary-Hamiltonian $\mathcal{ \tilde H}_1$ is able to drive the transition.

\noindent Secondly, the cutoff $M$ in Eq. (\ref{H1t}) implies approximating the function $f(k)$ in Eq. (\ref{fk}) by its truncated Fourier series. Since $f(k)$ is divergent and discontinuous at $g=g_c=1$ and $k=0$,  $\mathcal{\tilde H}_1(M)$ suffers from so-called Gibbs phenomenon,  this is, the problem of recovering point values of a nonperiodic or discontinuous function from its Fourier coefficients \cite{GS97}. In Fig. \ref{pkk}(a) we present the results for the truncation $\mathcal{\tilde H}_1(M)$ with $s_m=1$ (Dirichlet kernel).
The Gibbs phenomenon is seen here in the appearance of side-lobes at large $k$. This can be prevented by using a Fourier space filter $s_m$ that modifies the expansion coefficients \cite{GS97}.  In Fig \ref{pkk}(b) we use the raised cosine filter $s_m=\frac{1}{2}[1+\cos(m\pi/M)]$. 
It  improves the convergence away from the discontinuity, making the decay of  $p_k$ with $k$ (almost) monotonic, and suppresses the side-lobes observed in its absence at the expense of broadening $p_k$. However, it remains impossible to recover $f(k)$ close to its discontinuity,  so modes with $k \approx 0$ are still excited. 

\noindent As an upshot,  in the limit of  fast quenches the effects of approximating  $\mathcal{ H}_1(M)$ by  $\mathcal{\tilde H}_1(M)$ depends only on $M$, and to recover a fully adiabatic dynamics we need $M=N/2$. Based on the above considerations and the relation of $h_m$ to the correlation function in the Ising model, we can draw the conjecture that an approximation of the form in Eq. (\ref{H1t}) induces an adiabatic dynamics of the modes with $k \gg k_M \sim M^{-1}$. This is corroborated in the insets in Fig. \ref{pkk} where we rescale  $k$ for different values of the cutoff $M$, and the corresponding excitation probabilities 
$p_k$ collapse onto each other.

\begin{figure}
\begin{center}
\includegraphics{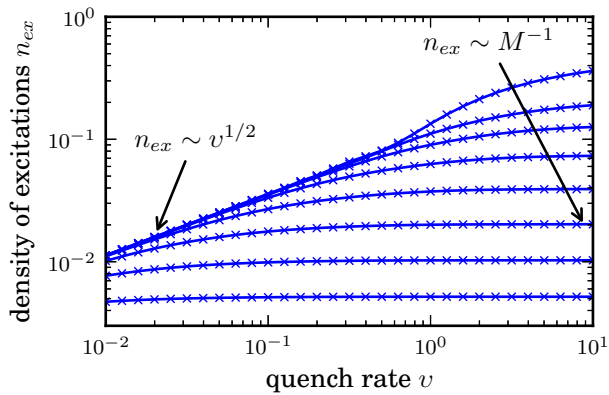}
\end{center}
\caption{\label{tqscaling} (Color online) Suppression of the total number of excitations $n_{\rm ex}$ as a function of the quench rate $\upsilon$ of the transverse field in a 1$d$ Ising chain
following a quench through the QCP ($g=1$). The dynamics is  assisted  by a truncated auxiliary Hamiltonian $\mathcal{\tilde H}_1(M)$, with a cutoff $M=0,1,2,4,8,16,32,64$ from top to bottom. 
The numerics for $\mathcal{H}_0$ ($m=0$) is the reference case where the KZM dictates a power-law scaling of $n_{\rm ex}$ for $\upsilon\ll 1$.
This scaling is recovered at slow quench rates in a  passage through the QCP assisted by $\mathcal{\tilde H}_1(M)$, while at faster rates there is an efficient suppression of excitations. As the range of the interactions is increased, the dynamics in all modes is driven through the instantaneous eigenbasis of $\mathcal{H}_0$, and a complete suppression of excitations is achieved.
$N=1600$, we evolve from $g_i=10$ to $g_f=0$, and no filtering is applied ($s_m=1$).
}
\end{figure}

\noindent We consider as well  the mean number of excitations, $n_{\rm ex}=\frac{1}{\pi}\int_0^{\pi}p_k dk$,  as a function of quench rate $\upsilon$ and  cutoff $M$. The results are presented in Fig. \ref{tqscaling}. In a fast transition and for a given cutoff $M$, we are able to adiabatically drive modes with $k \gg k_M \sim M^{-1}$. This means that the mean number of defects saturates at $n_{ex}\sim M^{-d}$. This limit can be seen at the right hand side of Fig. \ref{tqscaling}.

%

Next, we focus on slower transitions which are induced by the Ising Hamiltonian and approximated supplementary Hamiltonian $\mathcal{H} = \mathcal{H}_0+ \mathcal{\tilde H}_1 (M)$. In the limiting case of $M=0$ the system is driven only by the Ising Hamiltonian and the non-adiabatic dynamics is correctly described by KZM 
\cite{ZDZ05,Dziarmaga05}  -- we do not consider here the super-adiabatic limit where the quench across QCP is adiabatic due to the gap resulting from the finite size of the system, which is reached for extremely slow quenches with $\upsilon  \ll N^{-2}$ \cite{Murg04,Dziarmaga05}.

\noindent For $M=0$, there appears a characteristic value of momenta described by KZM: $k_{KZ}\sim \upsilon^{\nu/(1+z\nu)}=\upsilon^{1/2}$ for the Ising model \cite{Polkovnikov05} (to be precise, we expect the power-law behavior for $\upsilon\ll 1$, when the system goes out of equilibrium close to the QCP).  The modes with $k\gg k_{KZ}$ are expected to cross QCP adiabatically and $n_{ex} = \upsilon^{d \nu / (1+d \nu)}=\upsilon^{1/2}$. We recover this limit in the left hand side of Fig. \ref{tqscaling}, when $M$ is small enough compared to $\upsilon^{-1}$. 
A crossover between the two quench rate limits occurs for intermediate  values of $M$ and $\upsilon$.
The one which dominates for a given set of parameters depends on whether $k_{KZ}$ is smaller or greater them $k_M$.

{\it Relation to the fidelity susceptibility.-}
Finally, it is interesting to draw a connection with the so-called fidelity susceptibility, which puts some constrains on $\mathcal{H}_1$.
Fidelity susceptibility, $\chi_F(\lambda)$, can be defined in the leading order expansion of the overlap of the ground-states calculated for slightly different values of external parameter $\lambda$. For a finite system, in the limit $\delta \rightarrow 0$ \cite{fullfidelity}, we can Taylor expand the overlap in $\delta$ \cite{FS}:
$$
|\la \lambda | \lambda+\delta \ra|^2 \approx 1- \delta^2 \chi_F(\lambda).
$$
where $\chi_f(\lambda)$ for a non-degenerated ground state reads \cite{FSscaling3,Commentparity}
$$
\chi_f(\lambda) = \sum_{n\neq0} \frac{|\la 0(\lambda) |\partial_\lambda \mathcal{H}_0|n(\lambda)\ra|^2}{|E_n - E_0|^2}, 
$$
and $ \{ | n (\lambda) \ra \} $ and $E_n$ are instantaneous eigenstates and eigenenergies of $\mathcal{H}_0(\lambda)$, and $|0(\lambda)\ra$ is the ground-state.

In addition, the supplementary Hamiltonian $\mathcal{H}_1$ (\ref{H1full}) which would be able to drive the evolution along the instantaneous {\it ground state}  must satisfy 
\beqa 
 \la0 (\lambda) | \mathcal{H}_1 | n(\lambda)\ra=i \lambda'(t) \frac{\la 0(\lambda) |\partial_\lambda \mathcal{H}_0|n(\lambda)\ra}{E_n-E_0},
\eeqa
for $n\neq 0$. Thus, we can verify that the mean variance of $\mathcal{H}$ in the instantaneous ground state is \cite{CommentH}
\beqa
\label{varH1}
\Delta \mathcal{H}^2 = \la 0(\lambda) |\mathcal{H}^2_1| 0(\lambda)\ra=|\lambda'(t)|^2\chi_F(\lambda),
\eeqa
Fidelity susceptibility, for translationally invariant system, is expected to {\it typically} scale \cite{FSscaling1,FSscaling2,FSscaling3}  
at the critical point as $\chi_F(\lambda_c) \sim N^{2/d\nu}$, and away from the critical point as $\chi_F(\lambda) \sim N |\lambda-\lambda_c|^{d\nu-2}$.  
It is divergent in the vicinity of the QCP, as long as fidelity susceptibility is dominated by low lying excitations \cite{FSscaling3}.

{\it In conclusion}, for a broad family of many-body systems exhibiting a quantum phase transition, 
we have presented a method to assist the adiabatic crossing of the critical point at a finite-rate by supplementing the system with  a finite-range time-dependent interaction.
Our proposal is suited to access the ground state manifold in quantum simulators.
The non-local terms of  $\mathcal{H}^{[m]}_1$-type in the auxiliary Hamiltonian can be efficiently implemented using  the  stroboscopic techniques recently demonstrated in the laboratory \cite{kbody,Barreiro11,Casanova12}.
We have focused on the finite-rate adiabatic crossing of a quantum phase transition, 
where suppressing excitations is particularly challenging due to the critical slowing down in the proximity of the critical point. 
Nonetheless, the method can be applied as well to the preparation of many-body states as an alternative to optimal control techniques \cite{RC11} or in combination with them.

{\it Acknowledgment.-} 
Discussions with  J. Garc\'ia-Ripoll, D. Porras and F. Verstraete, as well as comments by B. Damski and M. B. Plenio are gratefully acknowledged. 
This research is supported by the U.S Department of Energy through the LANL/LDRD Program and a  LANL J. Robert Oppenheimer fellowship (AD).
MMR acknowledges support from the ERC grant Querg.


\begin{thebibliography}{99}
\bibitem{CommentH}  $H_1$ is the part of the Hamiltonian which is responsible for driving of the quantum state. 
The presence of $H_0$ only results in the global phase of the system, and can --  in principle -- be dropped. 
We use the convention where $\langle n | H_1| n \rangle = 0$.
\bibitem{Commentparity} To be precise, we are confined here in the subspace with plus-one-parity which e.g. in the ferromagnetic phase single out one  ground state which does not break the ($Z_2$) parity symmetry.
\end{thebibliography}
\end{document}